\documentclass[ twocolumn]{aastex62}

\usepackage{sidecap}
\usepackage{color}

\newcommand{\rosat}{{\it ROSAT  \,}}
\newcommand{\gaia}{{\it Gaia \,}}

\newcommand{\galex}{{\it GALEX  \,}}

\newcommand{\ms}{M$_{\odot}$}

\newcommand{\fluxcgs}{erg~s$^{-1}$~cm$^{-2}$}

\graphicspath{{./}{}}

\shorttitle{An Analysis of the UV and X-Ray Evolution of Low-Mass Stars in the Era of \textit{Gaia}}
\shortauthors{Richey-Yowell et al. (2023)}

\begin{document}
\title{HAZMAT. IX. An Analysis of the UV and X-Ray Evolution of Low-Mass Stars in the Era of \textit{Gaia}}

\author[0000-0003-1290-3621]{Tyler Richey-Yowell}
\affil{Lowell Observatory, Flagstaff, AZ 86004, USA}
\affil{Percival Lowell Postdoctoral Fellow}
\affil{School of Earth and Space Exploration, Arizona State University, Tempe, AZ 85281, USA}
\email{try@lowell.edu}

\author[0000-0002-7260-5821]{Evgenya L. Shkolnik}
\affil{School of Earth and Space Exploration, Arizona State University, Tempe, AZ 85281, USA}

\author[0000-0002-6294-5937]{Adam C. Schneider}
\affil{United States Naval Observatory, Flagstaff Station, 10391 West Naval Observatory Rd., Flagstaff, AZ 86005, USA}

\author[0000-0002-1046-025X]{Sarah Peacock}
\affil{NASA Goddard Space Flight Center, Greenbelt, MD 20771, USA}
\affil{University of Maryland, Baltimore County, Baltimore, MD 21250}

\author{Lori A. Huseby}
\affil{Lunar and Planetary Laboratory, University of Arizona, Tucson, AZ 85721, USA}

\author[0000-0003-0711-7992]{James A. G. Jackman}
\affil{School of Earth and Space Exploration, Arizona State University, Tempe, AZ 85281, USA}

\author[0000-0002-7129-3002]{Travis Barman}
\affil{Lunar and Planetary Laboratory, University of Arizona, Tucson, AZ 85721, USA}

\author{Ella Osby}
\affil{School of Earth and Space Exploration, Arizona State University, Tempe, AZ 85281, USA}

\author[0000-0002-1386-1710]{Victoria S. Meadows}
\affil{NASA Nexus for Exoplanet System Science, Virtual Planetary Laboratory, University of Washington, Seattle, WA 98195, USA}

\accepted{May 3, 2023}
% \revised{\today}
\submitjournal{ApJ}

\begin{abstract}
Low mass stars ($\leq 1$ \ms) are some of the best candidates for hosting planets with detectable life because of these stars' long lifetimes and relative planet to star mass and radius ratios. An important aspect of these stars to consider is the amount of ultraviolet (UV) and X-ray radiation incident on planets in the habitable zones due to the ability of UV and X-ray radiation to alter the chemistry and evolution of planetary atmospheres. In this work, we build on the results of the HAZMAT I \citep{Shkolnik2014} and HAZMAT III \citep{Schneider2018} M star studies to determine the intrinsic UV and X-ray flux evolution with age for M stars using \gaia parallactic distances. We then compare these results to the intrinsic fluxes of K stars adapted from HAZMAT V \citep{richey-yowell2019}.  We find that although the intrinsic M star UV flux is 10 to 100 times lower than that of K stars, the UV fluxes in their respective habitable zone are similar. However, the habitable zone X-ray flux evolutions are slightly more distinguishable with a factor of 3 -- 15 times larger X-ray flux for late-M stars than for K stars. These results suggest that there may not be a K dwarf advantage compared to M stars in the UV, but one may still exist in the X-ray.
    
\end{abstract}

\section{Introduction}\label{sec:intro}

There are many factors that go into determining planet habitability, one of which is the understanding the luminosity evolution of the star \citep[e.g.][]{Meadows2018}. The high-energy radiation environment, i.e. the ultraviolet (UV) and X-ray radiation, of host stars can severely impact planetary atmospheres, as these wavelengths can cause photo-dissociation and ionization of molecules and atoms in the atmosphere, leading to changes in the composition or even complete erosion of the atmosphere \citep[e.g.][]{Segura2005, Airapetian2017, Tilley2019, Teal2022}. Additionally, this photochemistry can produce hazes in the atmospheres of planets that make observations of any molecular signatures quite difficult \citep{Zerkle2012, Neves2022a}. Understanding the high-energy radiation environment of low-mass stars and how this evolves in time is thus critical for modeling planetary atmosphere composition and being able to observe potential biosignatures of planets.

The HAbitable Zones and M star Activity across Time (HAZMAT) program aims to determine the temporal UV and X-ray evolution of low-mass stars with the goal of providing the necessary constraints for habitability studies of planets around these types of stars. The low-mass stars are divided into three luminosity regimes: K stars (0.6 -- 0.9 \ms), partially-convective early-type M stars (M1-3, 0.6 -- 0.35 \ms), and fully-convective late-type M stars (M4-9, 0.08–0.35 \ms). In this paper, we build on the HAZMAT I \citep[][henceforth SB14]{Shkolnik2014}, HAZMAT III \citep[][henceforth SS18]{Schneider2018}, and HAZMAT V \citep[][henceforth RY19]{richey-yowell2019} works, which utilized J-band normalization where distances were unavailable to analyze the evolution of UV/J and X-ray/J flux density ratios. In this paper, we analyze the evolution of the intrinsic fluxes using recently released \gaia data for the same set of stars and compare the distance-corrected and distance-independent approaches. These previous papers utilized archived \galex photometry of stars which are confirmed members of young moving groups (see \S 2.1) to analyze the UV/J evolution of M and K stars as a function of age:

\textit{SB14.} The results of SB14 demonstrated that the near-UV (NUV, 1771 -- 2831\AA),  far-UV (FUV, 1344 -- 1786\AA), and X-ray (0.6 -- 300 \AA) flux density ratios of early-M stars remain constant for 300 Myr of the stars' lifetimes before beginning to decline. Both the UV and X-ray showed two distinct groups of emitters -- those with high-activity and those with low-activity. 

\textit{SS18.} The results of SS18 showed that early- and late-type M stars experience different evolutionary trends.  For early-M stars, the UV/J flux density ratio for both NUV and FUV declines with age by a factor of 4. For the late-M stars, the NUV/J flux density ratio declines by a factor of 11, while the FUV/J flux density ratio declines by a factor of 31. These differences are attributed to the slower rotational evolution of late-M stars \citep{Newton2017}. 

\textit{RY19.} The results of RY19 showed that K stars experience similar levels of UV/J flux density ratio compared to M stars, within a factor of 10 between both early- and late-type M stars. With their farther and wider habitable zone distances from the star, planets in the habitable zones of K stars must therefore receive less UV flux than their M star counterparts, thus placing these planets at a potential advantage compared to those around M stars. 

Additionally, recent studies have shown a rotational stalling effect where the rotational evolution of K stars appears to be stagnant \citep{Curtis2019, Curtis2020}. This effect has several consequences for the potential habitability of planets, namely an elongated period of saturated UV activity as observed in a spectroscopic UV study of K stars \citep[][]{Richey-Yowell2022}. 

This work updates the results of SB14, SS18, and RY19 with recent distance measurements from \gaia, yielding  measurements of the intrinsic UV and X-ray flux evolutions of M stars to compare with those of K stars. We summarize the M and K star samples, \galex photometry, and \rosat photometry in \S\ref{sec:methods}. In \S\ref{sec:evolution}, we analyze the intrinsic M star activity evolution, followed by comparisons with K star data in \S\ref{sec:hz}. Finally, we discuss the differences in the results between the distance-independent and distance-corrected methods and their causes in \S\ref{sec:comparison}.

\begin{figure*}[t]
    \centering
    \includegraphics[width=\linewidth]{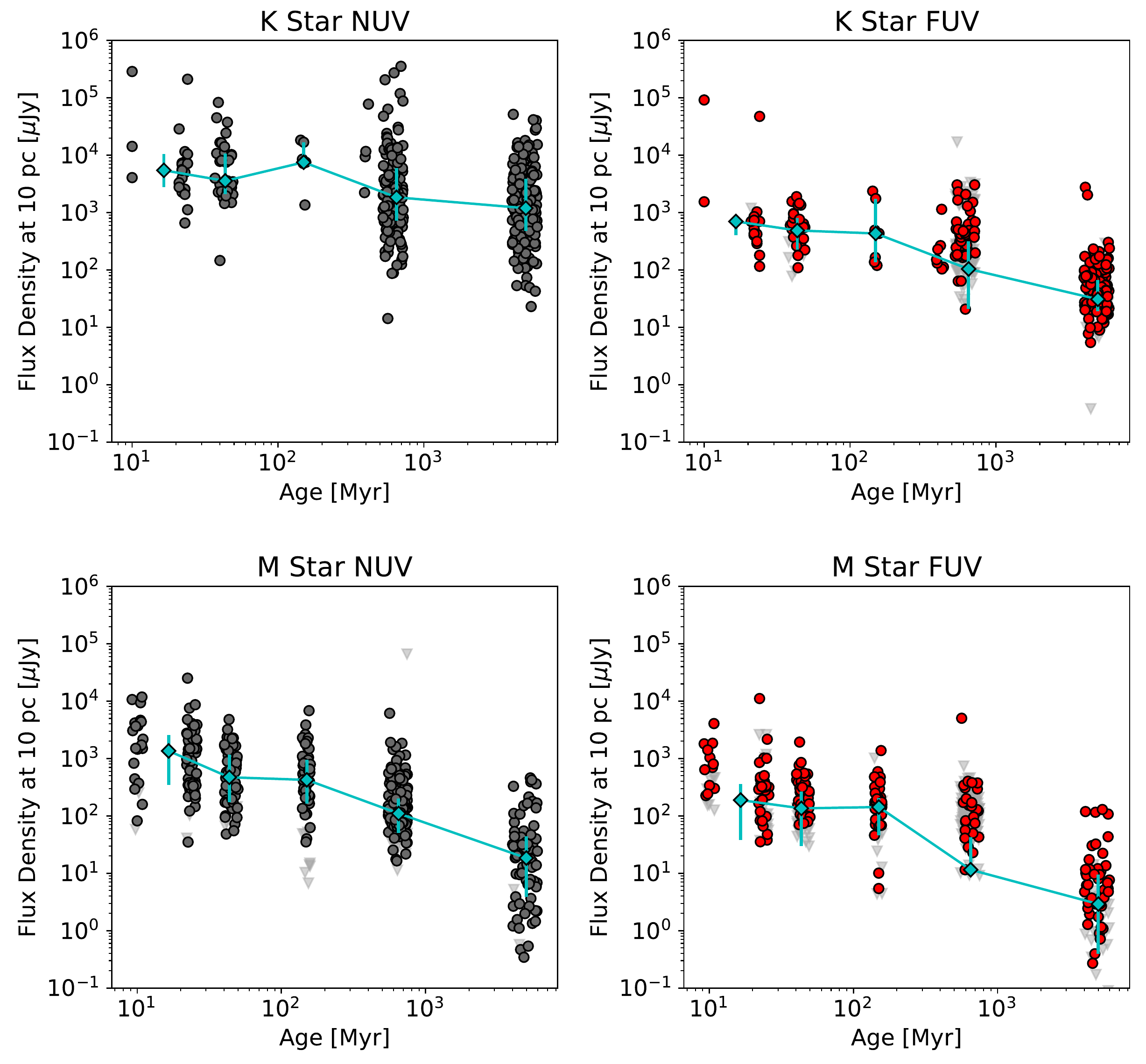}
    \singlespace\caption{The UV evolution of K (top) and M (bottom) stars with stellar age. The plots show the NUV (left) and FUV (right) intrinsic flux density at a standard distance of 10 pc. The circles are observations while the triangles represent upper limits. Typical flux density uncertainty is smaller than the marker. The median flux density values at each age bin are shown with teal diamonds and the error bars represent the interquartiles (i.e. the middle 50\%) of the data set, accounting for the upper limits. Note that the median for the M star 650 Myr sample is weighted significantly by the large number of upper limit values.}
    \label{fig:fluxevolution}
\end{figure*}

\begin{figure}[t]
    \centering
    \includegraphics[width=0.47\textwidth]{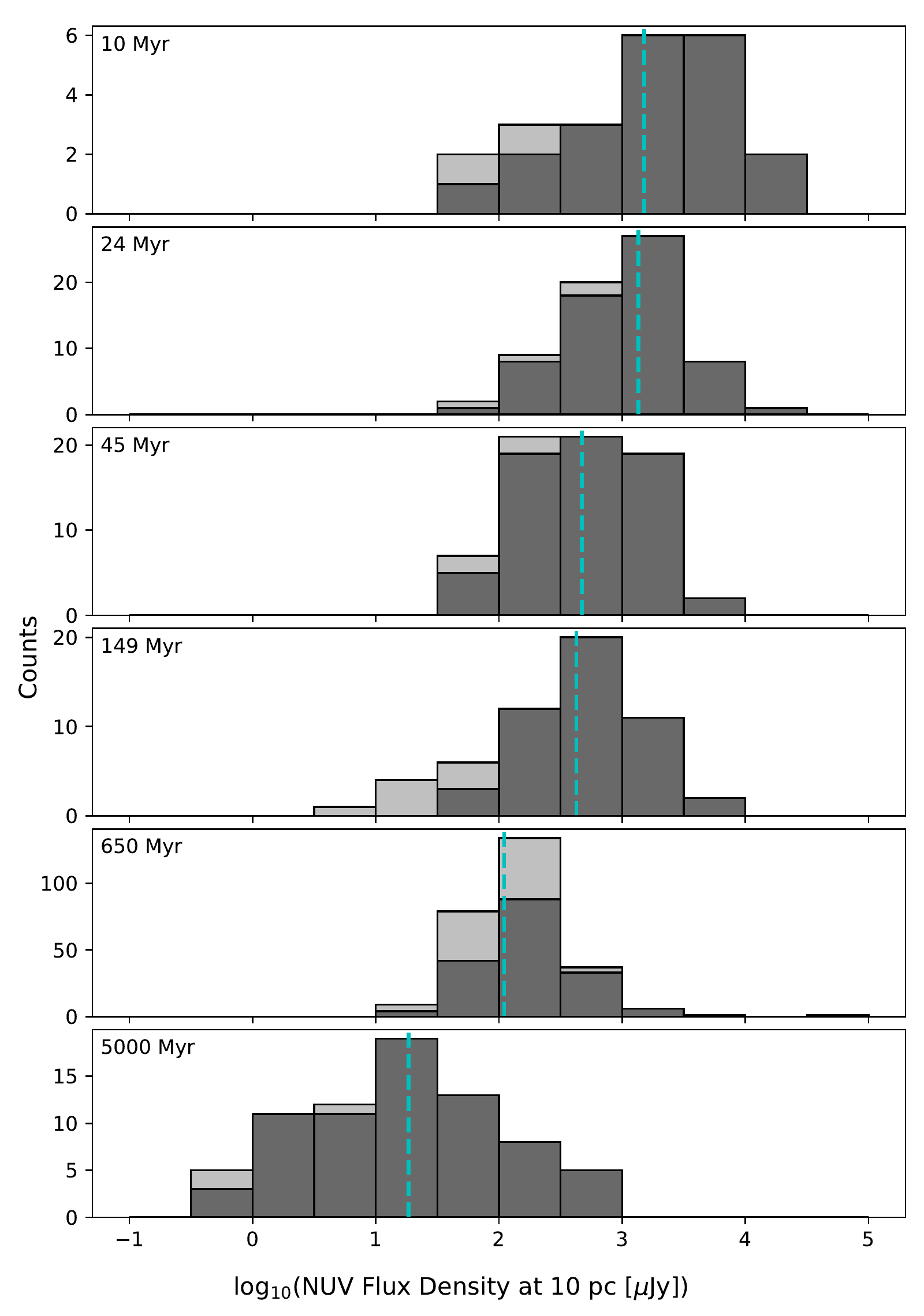}
    \singlespace\caption{Histogram of the intrinsic NUV flux density of M stars. Dark gray bars are detections, while light gray bars are upper limits. The median flux density is calculated using the Kaplan-Meier survival estimate to include the data limits (see text for details) and is represented by the dotted vertical line.}
    \label{fig:hist_nuv}
\end{figure}

\begin{figure}[t]
    \centering
    \includegraphics[width=0.47\textwidth]{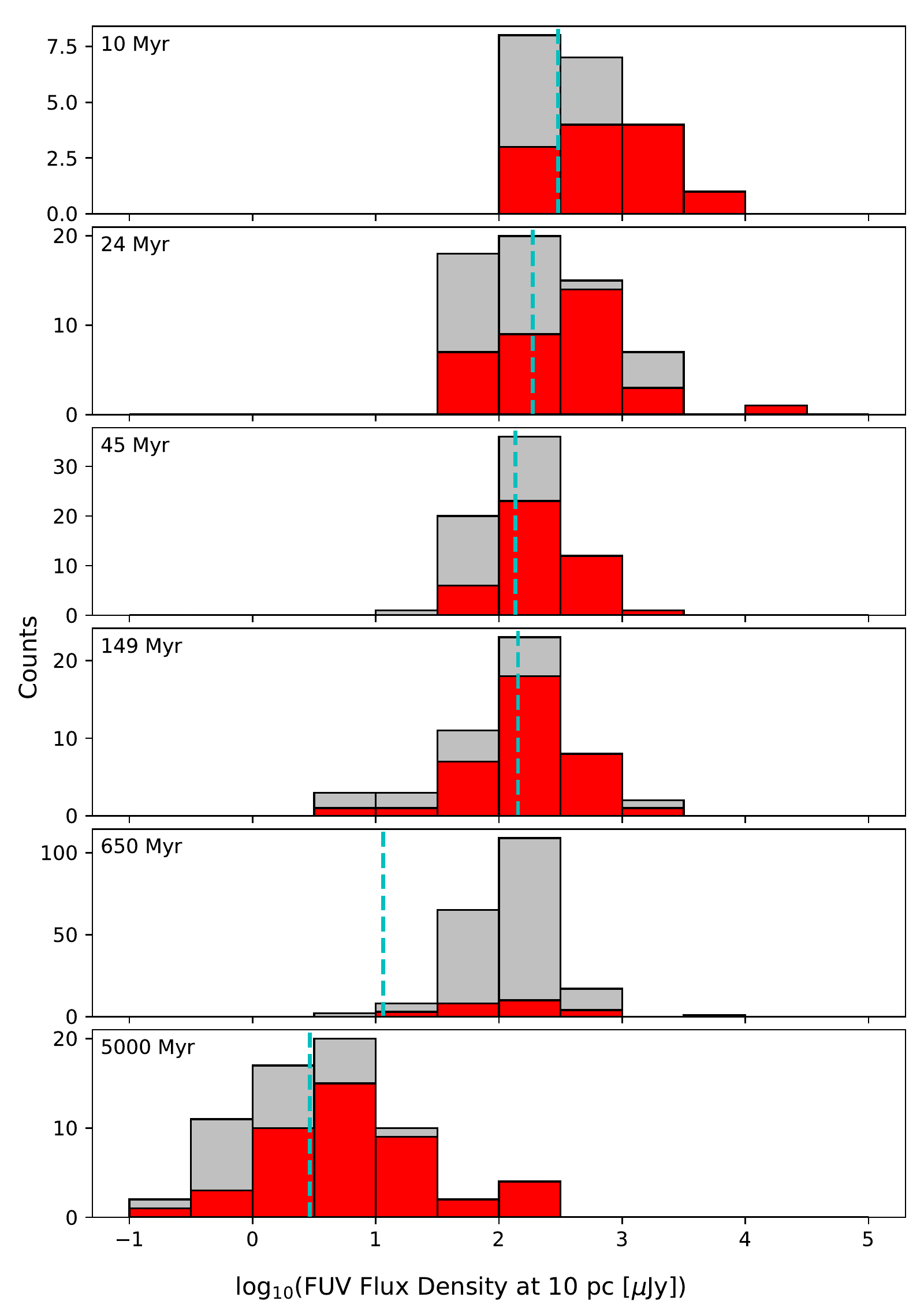}
    \singlespace\caption{Same as Figure \ref{fig:hist_nuv} but for the FUV flux density. Note that the median for the 650 Myr sample is weighted significantly by the large number of upper limit values. }
    \label{fig:hist_fuv}
\end{figure}

\begin{figure*}[t]
    \centering
    \includegraphics[width=\linewidth]{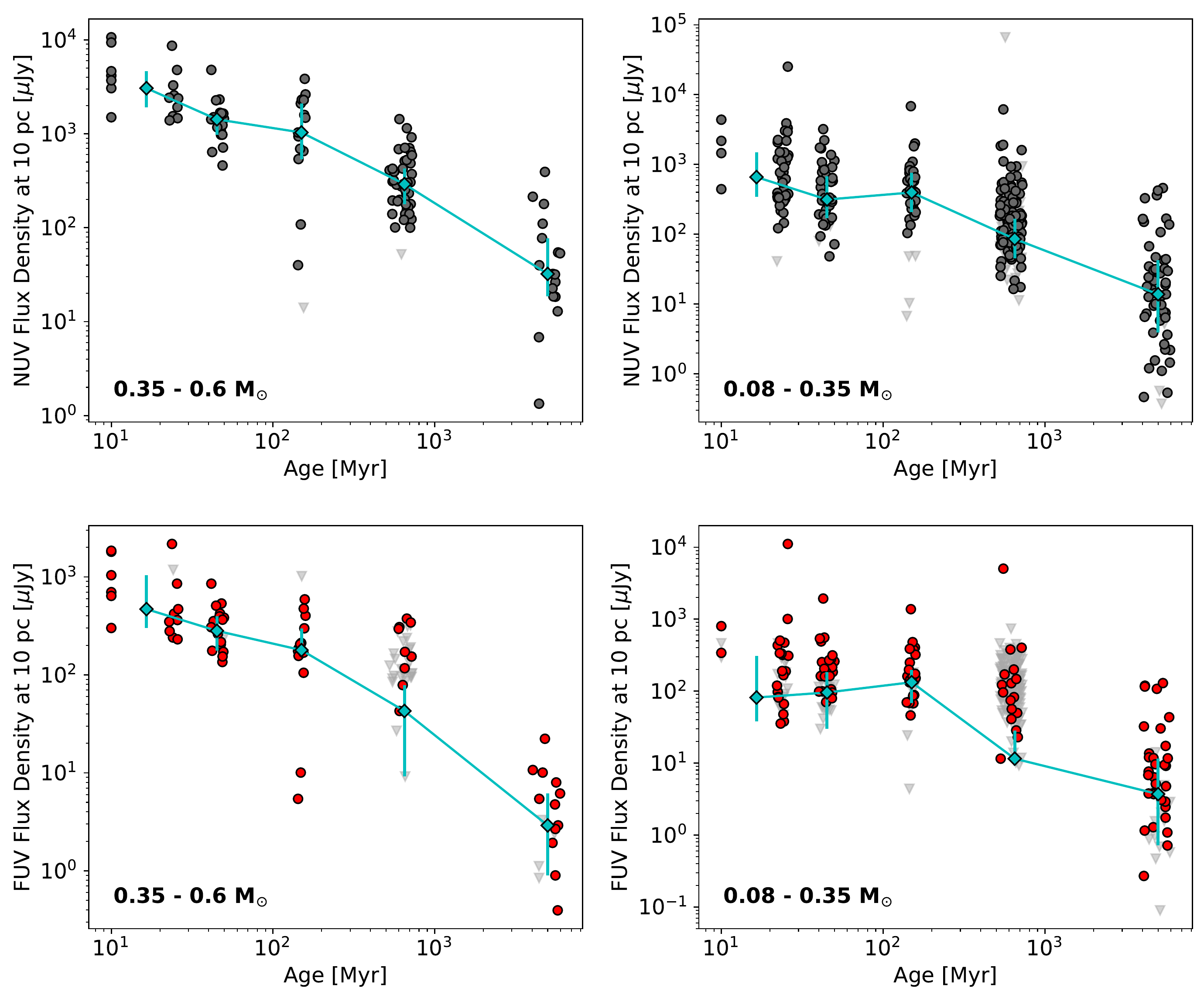}
    \singlespace\caption{The UV evolution of M stars, separated by early- (0.35 -- 0.6 \ms, left plots) and late-type (0.08 -- 0.35 \ms, right plots) M stars. The top and bottom plots show the NUV and FUV intrinsic flux density at a standard distance of 10 pc, respectively. The circles are observations while the triangles represent limits. Typical flux density uncertainty is smaller than the marker. The median flux density values at each age bin are shown with teal diamonds and the error bars represent the interquartiles (i.e. the middle 50\%) of the data set, accounting for upper limits. Note that the median for the 650 Myr sample is weighted significantly by the large number of upper limit values. There is a  distinction in the UV evolution between early- and late-type M dwarfs, with early-M stars evolving more rapidly, as also demonstrated using fractional flux densities by SS18.}
    \label{fig:fluxevolution_masses}
\end{figure*}

\begin{figure*}[t]
    \centering
    \includegraphics[width=0.7\linewidth]{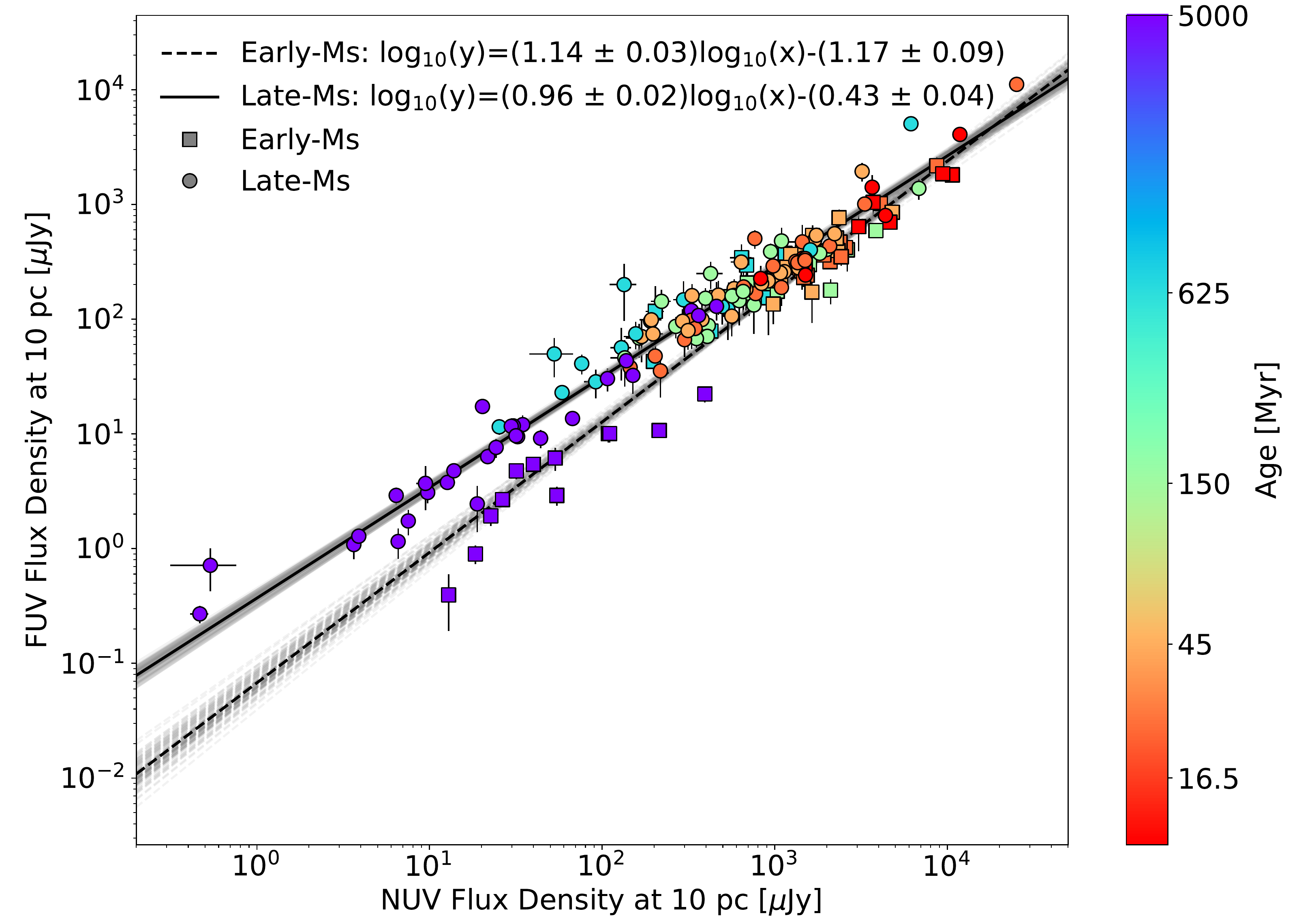}
    \singlespace\caption{M star absolute FUV flux density compared to absolute NUV flux density for stars with both FUV and NUV \galex detections. The color represents the age of the star. Early-M stars are squares and late-M stars are circles. The young stars remain clustered at similar FUV and NUV flux densities; however, the field-age stars show much lower emission. The separation between the two samples at the field age show the greater decline of the early-M star FUV flux compared to late-M stars.}
    \label{fig:fuv_v_nuv}
\end{figure*}

\begin{figure*}[t]
    \centering
    \includegraphics[width=\linewidth]{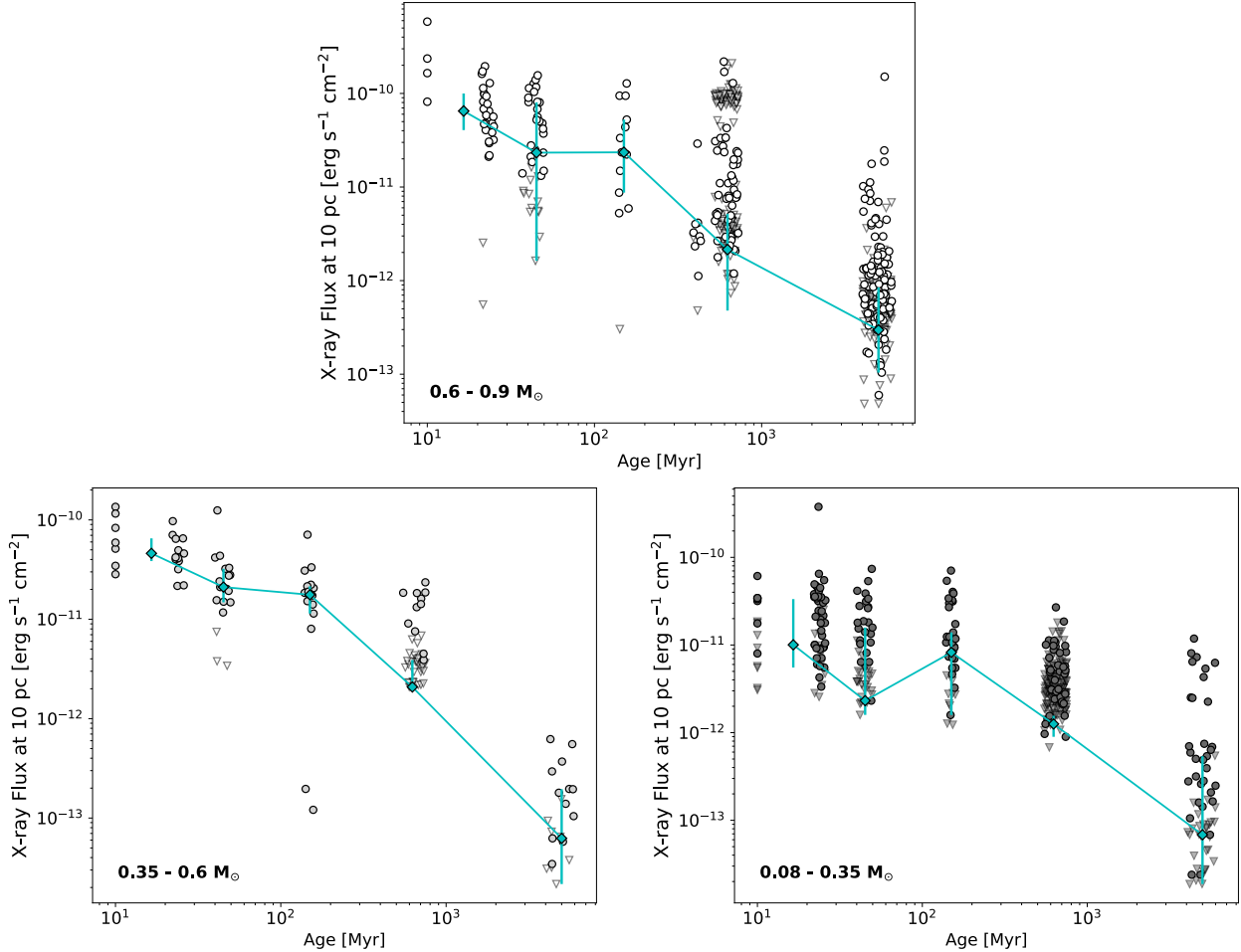}
    \singlespace\caption{The X-ray evolution of K and M stars as a function of stellar age, broken down by mass. The circles are \rosat detections while the triangles represent upper limits.  Typical flux uncertainty is smaller than the marker. The teal diamonds are the median values of the data in each age bin and the error bars represent interquartiles (i.e. the middle 50\% of the data), accounting for upper limits. Note that the medians are weighted significantly by the large number of upper limit values. The rates of decline are much steeper than those of the UV data.}
    \label{fig:xrayevolution}
\end{figure*}

\begin{figure}[t]
    \centering
    \includegraphics[width=0.47\textwidth]{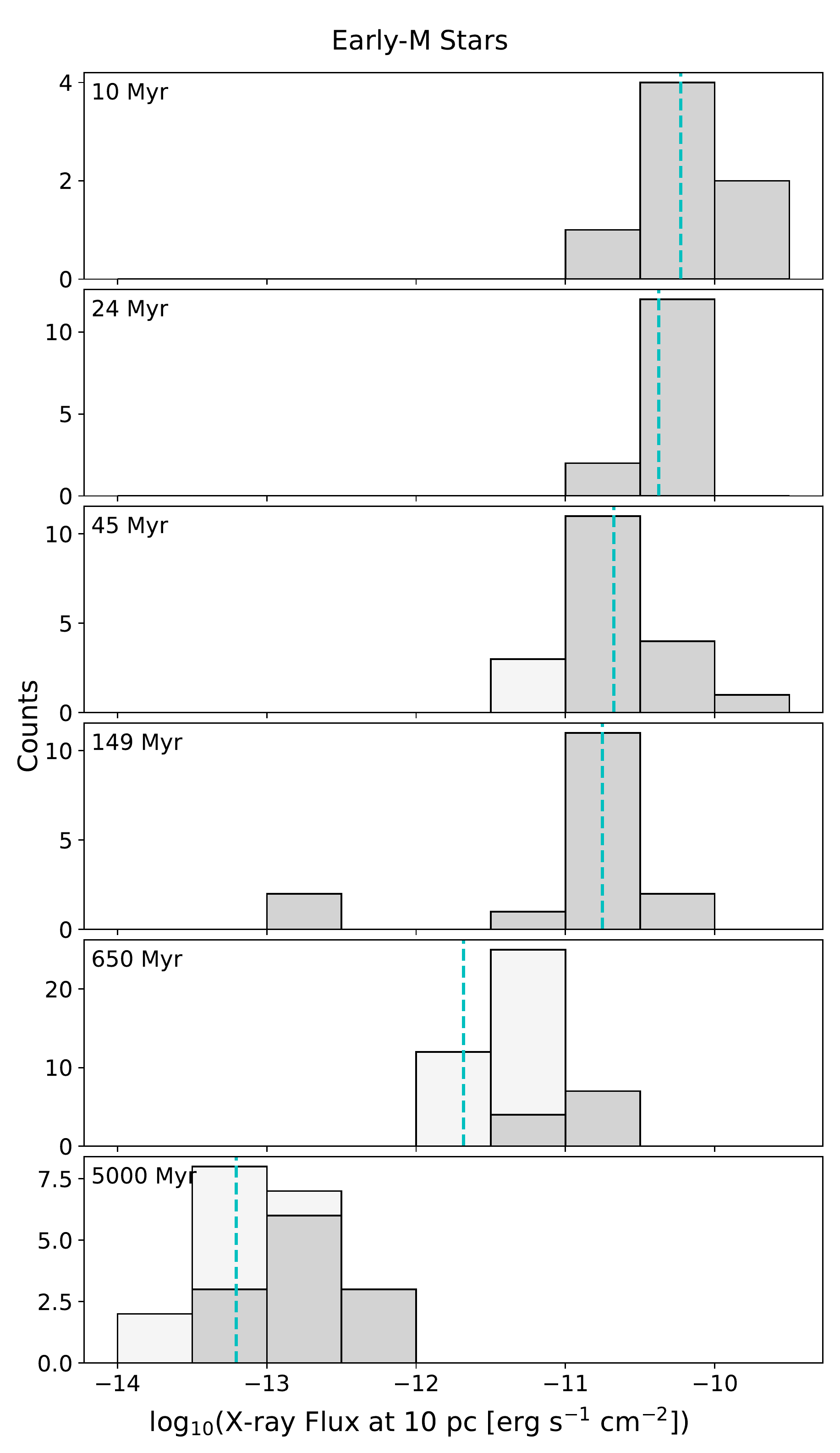}
    \singlespace\caption{Same as Figure \ref{fig:hist_nuv} but for X-ray Flux for Early-M stars. The darker color represents detections while the lighter color represents limits. }
    \label{fig:hist_xray_em}
\end{figure}

\begin{figure}[t]
    \centering
    \includegraphics[width=0.47\textwidth]{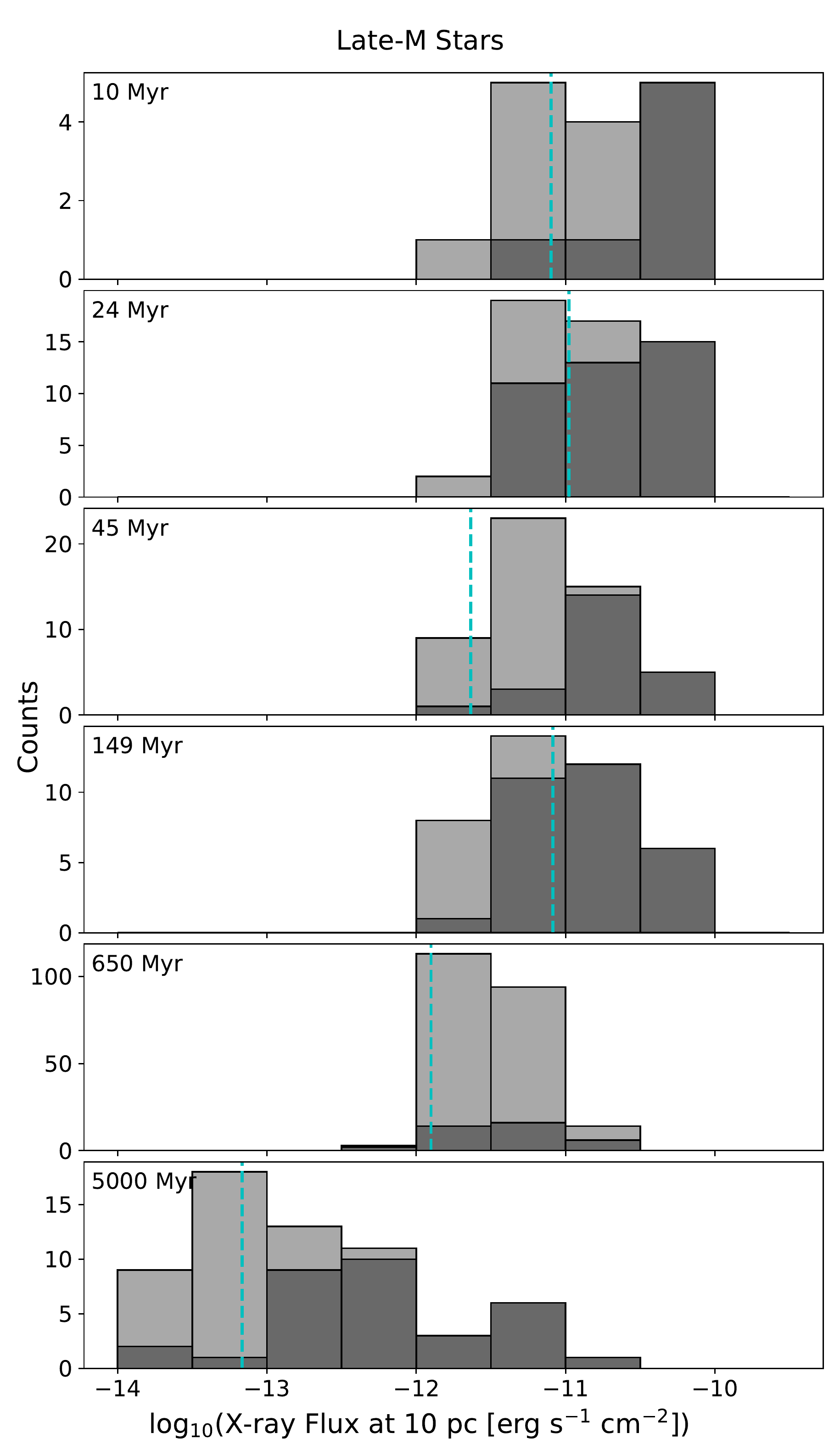}
    \singlespace\caption{Same as Figure \ref{fig:hist_nuv} but for X-ray Flux for Late-M stars. The darker color represents detections while the lighter color represents limits. }
    \label{fig:hist_xray_lm}
\end{figure}

\begin{figure}[t]
    \centering
    \includegraphics[width=0.47\textwidth]{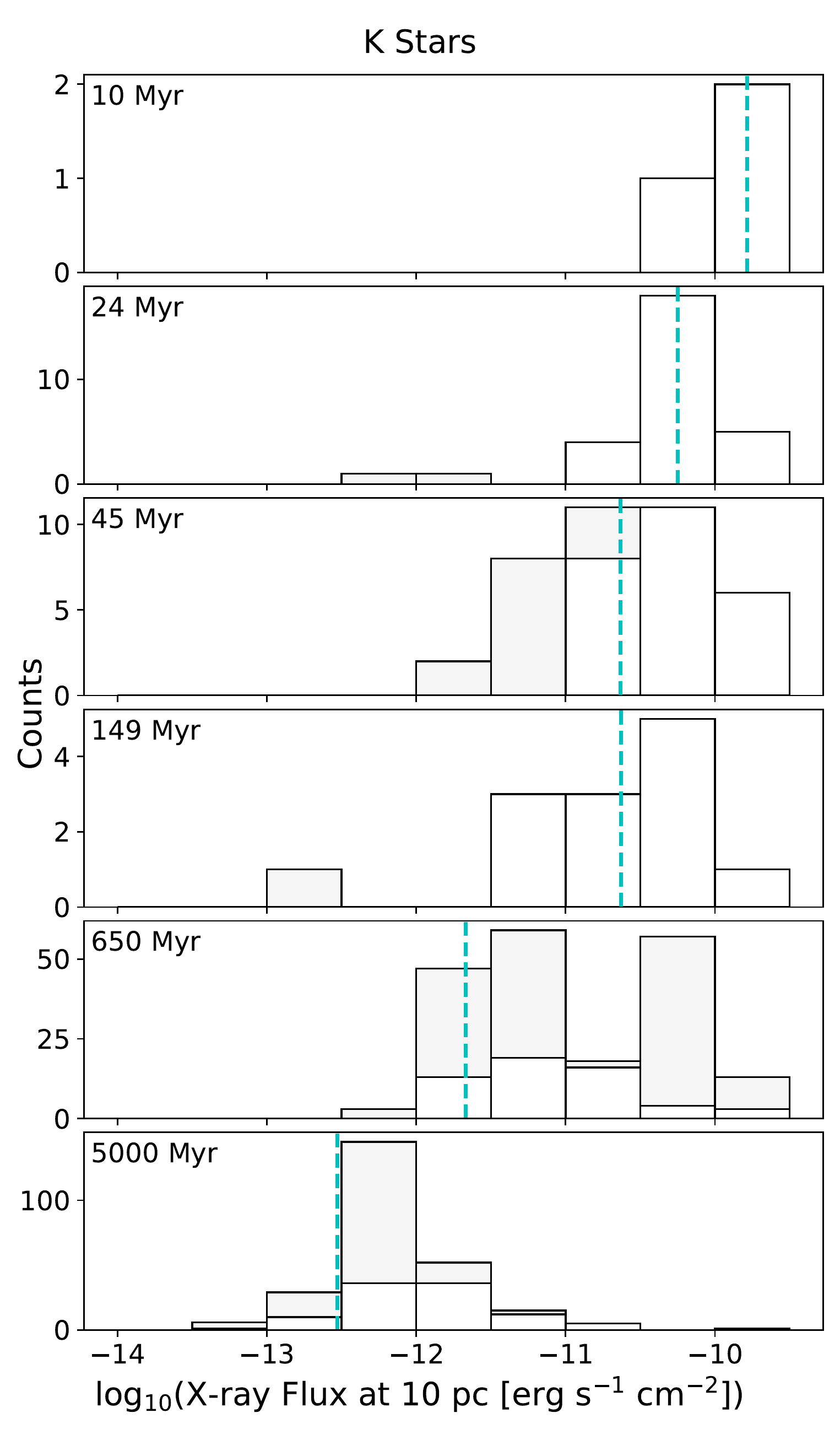}
    \singlespace\caption{Same as Figure \ref{fig:hist_nuv} but for X-ray Flux for K stars. The white represents detections while the gray represents limits. }
    \label{fig:hist_xray_k}
\end{figure}

% \begin{figure*}[t]
%     \centering
%     \includegraphics[width=0.7\linewidth]{uv_v_xray.pdf}
%     \singlespace\caption{The UV flux compared to the X-ray flux of M stars. The gray points are NUV detections while the red points are FUV detections. There is similar grouping of high- and low-activity stars as seen in the early-M stars in SB14 that does not appear to be correlated with spectral type.}. 
%     \label{fig:uv_vs_xray}
% \end{figure*}

\begin{figure*}[ht]
\centering
\gridline{\fig{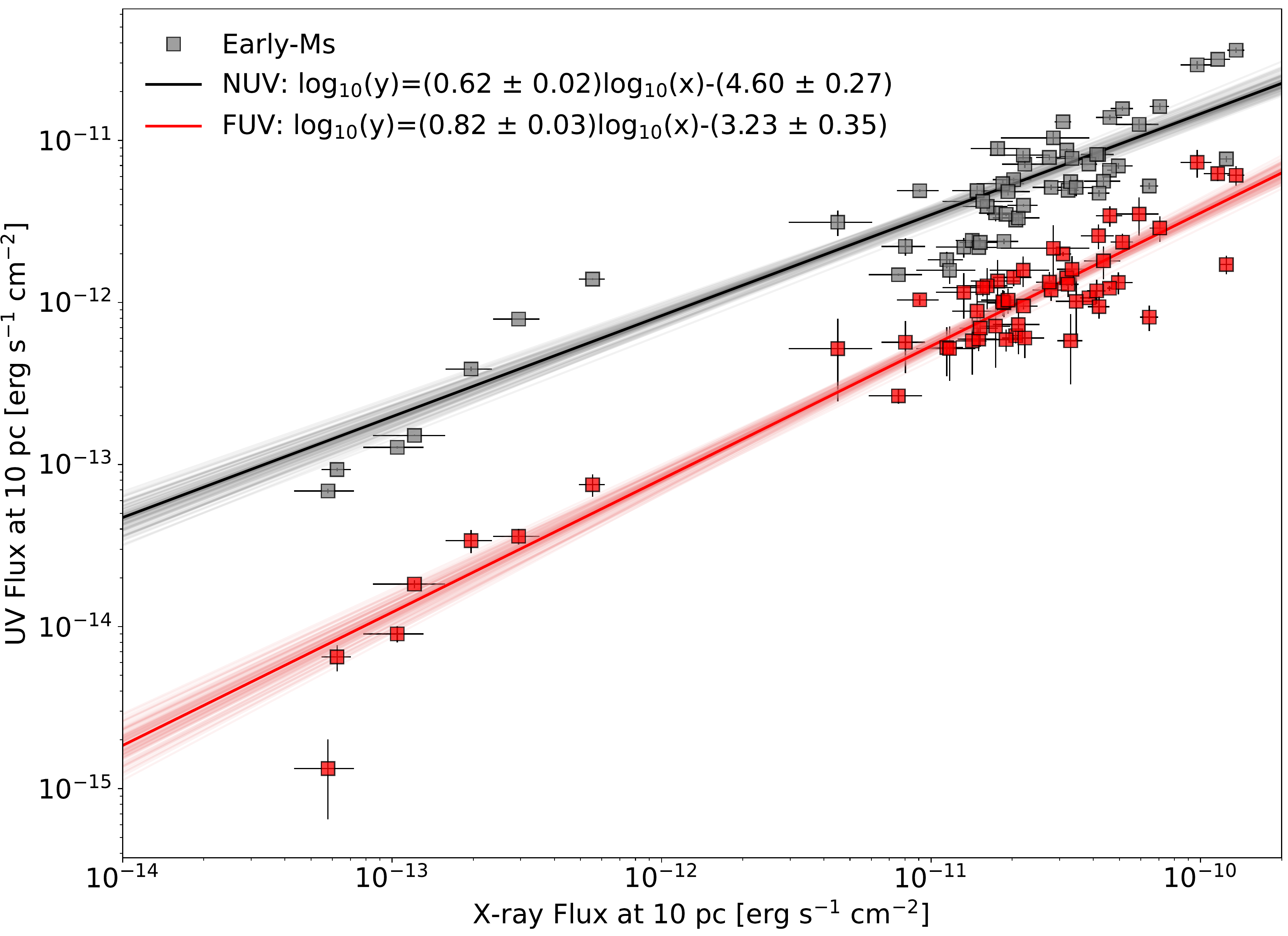}{0.5\textwidth}{(a)}
          \fig{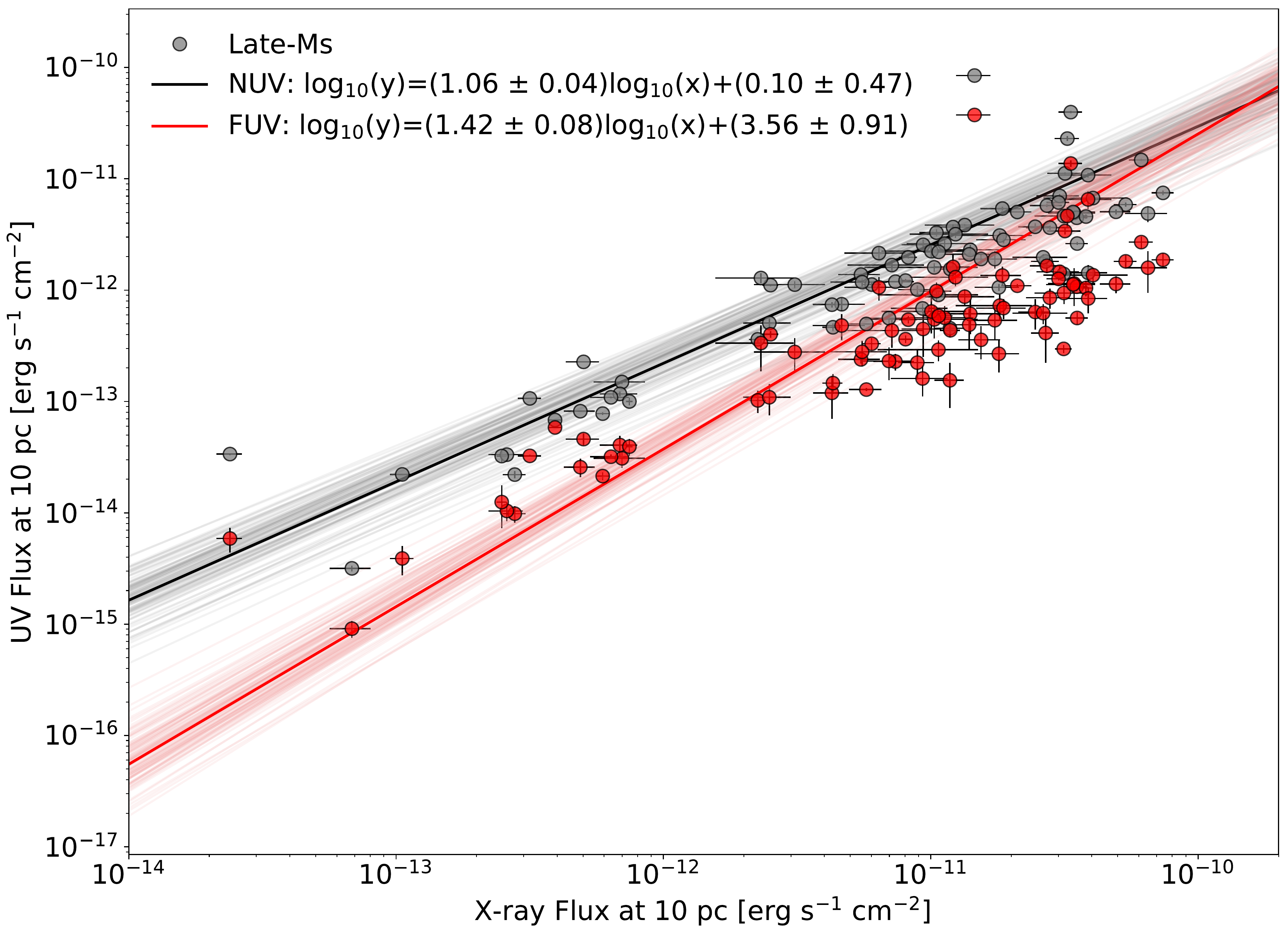}{0.5\textwidth}{(b)}
          }
\caption{The UV flux compared to the X-ray flux of early- (a) and late-type (b) M stars for stars with both UV and X-ray detections. The gray points are NUV detections while the red points are FUV detections. There is similar grouping of high- and low-activity stars as seen in the early-M stars in SB14 that does not appear to be correlated with spectral type.}
\label{fig:uv_vs_xray}
\end{figure*}

\begin{figure*}[t]
    \centering
    \includegraphics[width=0.9\linewidth]{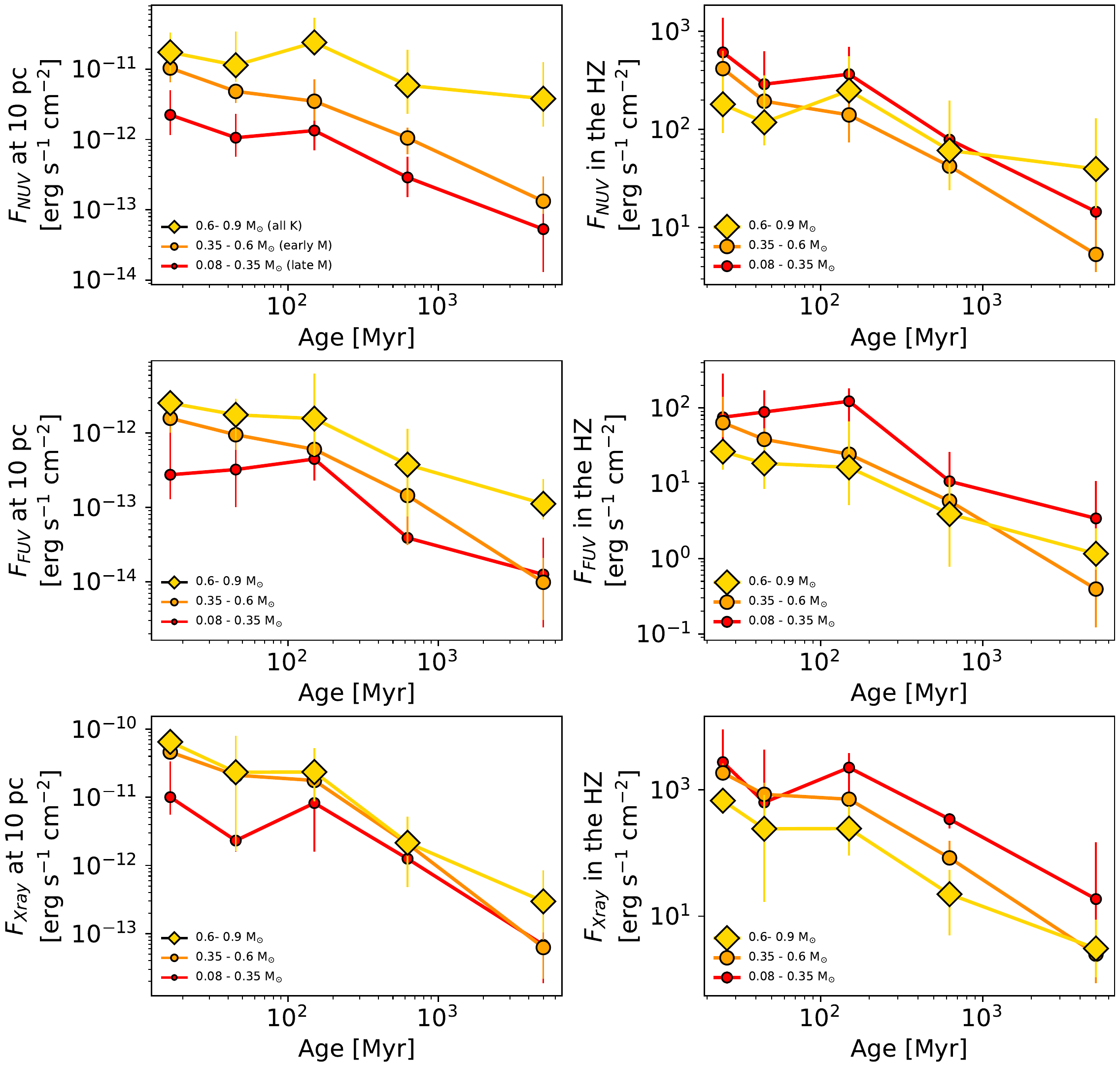}
    \singlespace\caption {Comparison of the NUV, FUV, and X-ray intrinsic and HZ median fluxes from K (yellow), early-M (orange), and late-M stars (red) in our sample. The medians and interquartiles are the same as Figures \ref{fig:fluxevolution_masses}, and  \ref{fig:xrayevolution} but converted to flux.}
    \label{fig:full_nuv_fuv}
\end{figure*}

% \begin{figure*}[t]
%     \centering
%     \subfloat[\centering a]{{\includegraphics[width=0.45\linewidth]{k_uv_v_j.pdf}}}
%     \qquad
%     \subfloat[\centering b]{{\includegraphics[width=0.45\linewidth]{m_uv_v_j.pdf} }}
%     \singlespace\caption[Comparison of the Distance-Corrected and Distance-Independent Approaches for Measuring UV Fluxes for K Stars and M Stars. ]{Comparison of the distance-corrected and distance-independent approaches for measuring UV flux densities for [a] K stars and [b] M stars. Red points are FUV data and black points are NUV data. Lines of best fit are overlaid and their equations shown in the labels. }
%     \label{fig:comps}
% \end{figure*}

\begin{figure*}[ht]
\centering
\gridline{\fig{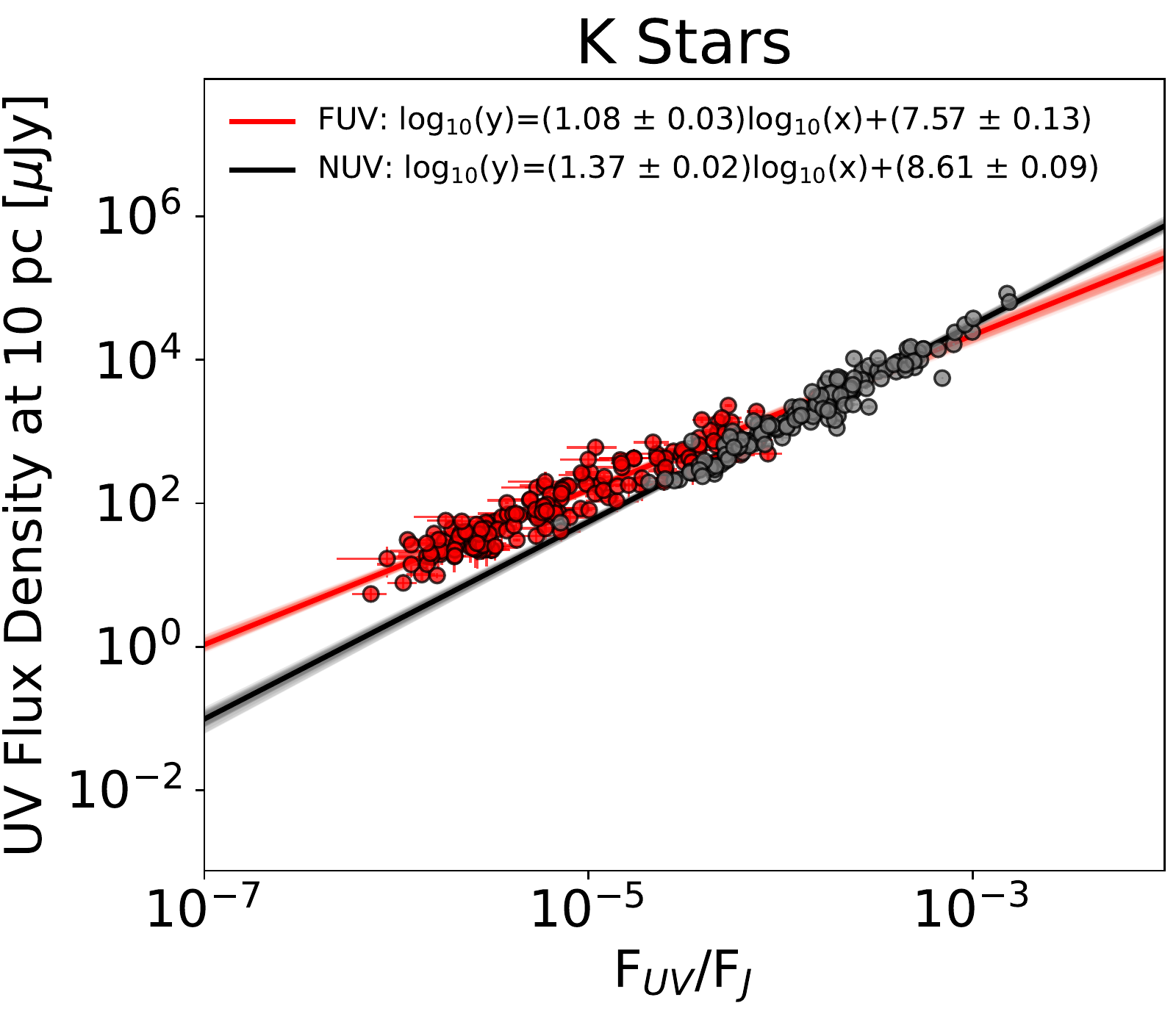}{0.5\textwidth}{(a)}
          \fig{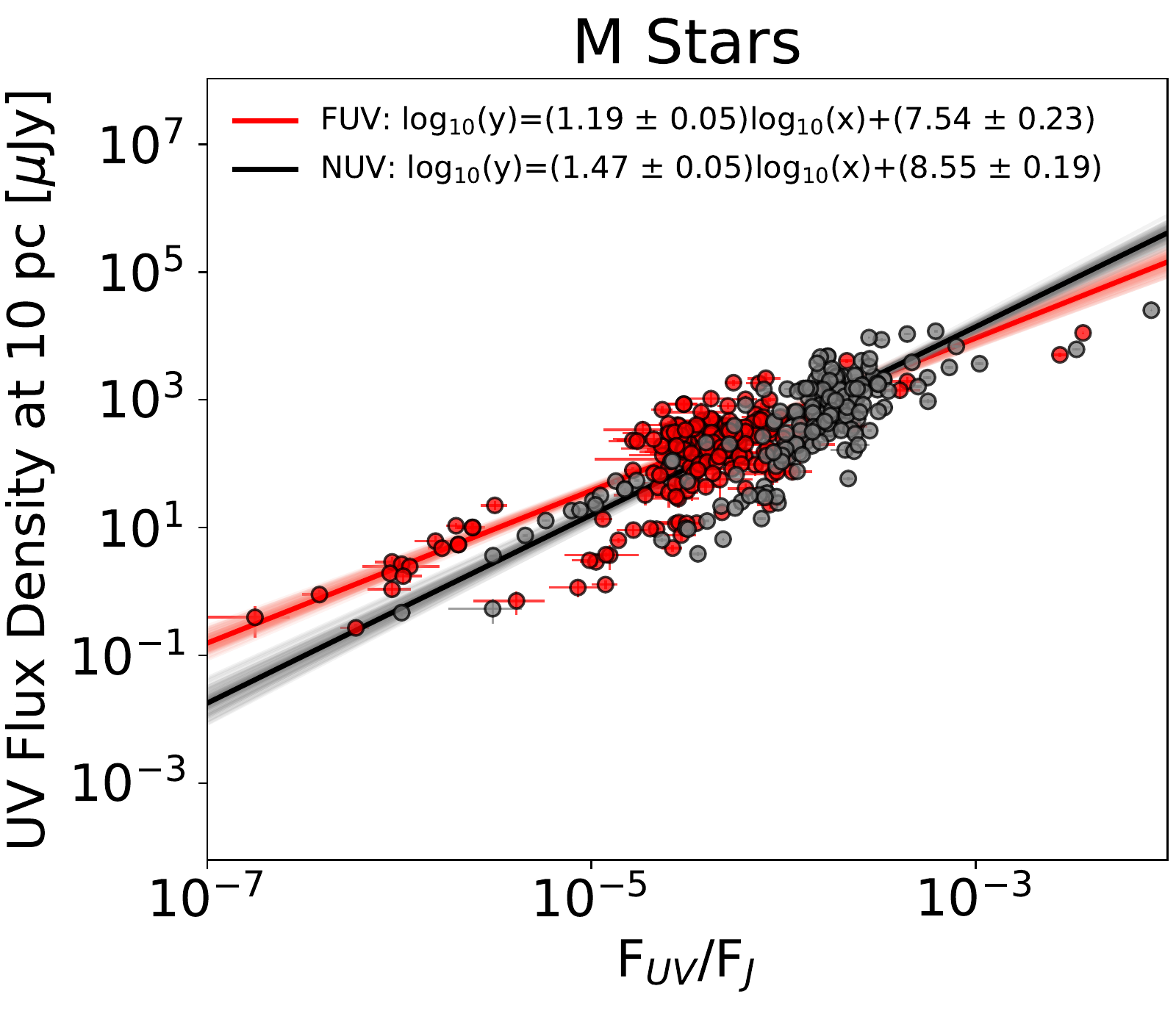}{0.5\textwidth}{(b)}
          }
\caption{Comparison of the distance-corrected and distance-independent approaches for measuring UV flux densities for (a) K stars and (b) M stars. Red points are FUV data and black points are NUV data. Lines of best fit are overlaid and their equations shown in the labels. }
\label{fig:comps}
\end{figure*}

\begin{figure}[t]
    \centering
    \includegraphics[width=\linewidth]{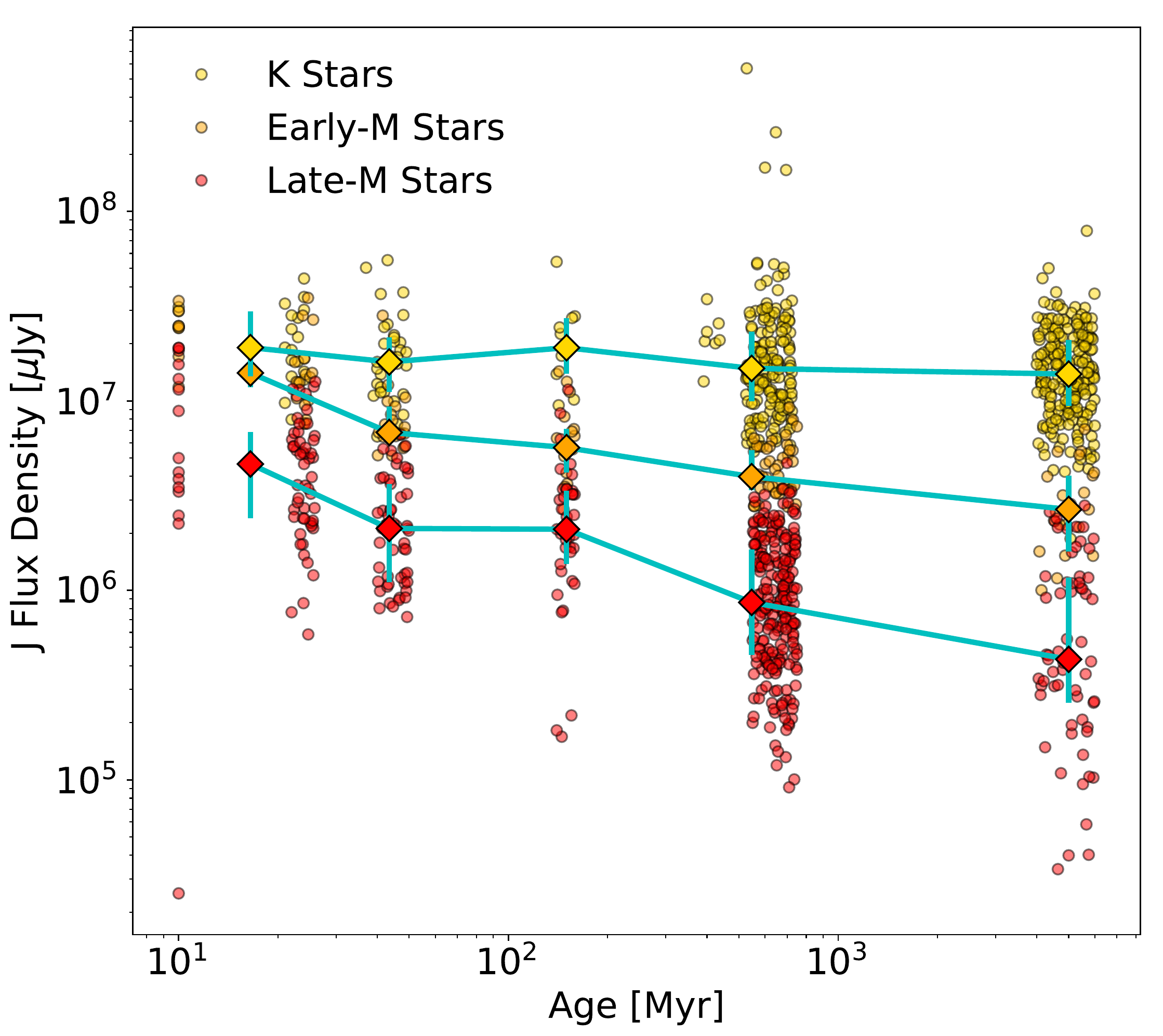}
    \singlespace\caption{The evolution of the distance-corrected J-band flux density for K and M stars. The evolution of the K star J-band flux density is flat, while the evolution for M stars decreases with age.}. 
    \label{fig:jbandevolution}
\end{figure}

\section{Target Selection and Archived Data Collection Methods}\label{sec:methods}

\subsection{Summary of the M and K Star Samples}

We utilize the same M and K star samples determined in SS18 and RY19. We refer the reader to these previous works for details, but summarize the procedures below:

We identified M and K star members of young moving groups (YMGs) and clusters ranging in age from 10 Myr to 650 Myr: TW Hydra (10 $\pm$ 3 Myr, \citealt{Reza1989, Bell2015}), Carina (13.3$^{+1.1}_{-0.6}$ Myr, \citealt{Torres2008, Booth2021}), Beta Pictoris (24 $\pm$ 3 Myr, \citealt{Zuckerman2001, Shkolnik2017}), Columba (42$^{+6}_{-4}$ Myr, \citealt{Torres2008, Bell2015}), Tucana-Horologium (45 $\pm$ 4 Myr, \citealt{KrausTucHor, Bell2015}), AB Doradus (149$^{+51}_{-19}$ Myr, \citealt{Zuckerman2004, Bell2015}), Ursa Majoris (414 $\pm$ 23 Myr;  \citealt{Eggen1992, Jones15b, Jones17}), Praesepe (600 Myr, \citealt{Crawford1969, Kraus2007}), and Hyades (650 $\pm$ 70 Myr, \citealt{Perryman1997, Martin2018}). We additionally ran a search in SIMBAD for M and K stars within 30 pc from Earth that were not identified as members of any known moving group or close-in binaries, to which we assigned an age of 5 Gyr, as this is the average age of the field stars nearby to our Sun. This yielded a total of 713 K stars and 977 M stars as the input sample. In this work, we remove two M stars and six K stars from this sample which have since been resolved as binaries in \gaia. We do not predict additional binary members in the YMGs due to the extensive spectrocopic observations carried out both in confirming membership within YMGs and in follow-up studies. In order to assess the potential for our field targets to be unknown binaries, we calculated the standard deviation of all of the high-quality radial velocity measurements of these stars, as a high radial velocity standard deviation would suggest orbital changes in the system consistent with a close-in binary system. We find only three stars with radial velocity standard deviations greater than 1 km/s, therefore the contribution by unknown binaries should be limited within our data set.

To compare the stellar samples by mass, we need to estimate the masses of our stellar sample. Starting with the spectral type of each star, we estimated the stellar effective temperature from the age-dependent relations from \citet{Pecaut2013}. We then converted from stellar effective temperatures into mass and radius estimates using the isochrone models from \citet{Baraffe2015}, which are a function of age. 

To accommodate the lack of measured parallaxes, SS18 normalized the stellar flux densities to the J band magnitude to yield age-independent normalized flux density ratios. With \gaia data \citep{gaiadr3} now available, we are able to measure the intrinsic stellar UV and X-ray fluxes instead. To get the parallactic distances, we searched the \gaia archive with a radius of 5\arcsec. For any stars with multiple objects within that radius, we verified which object was our target by a combination of separation from our input coordinates and confirming the other \gaia measurements were within normal ranges for M and K stars for our target. 

\subsection{Summary of the \galex Photometry}\label{sec:photometry}

We use the M and K star UV flux densities reported in SS18 and RY19. We refer the reader to these previous works for details, but summarize the procedures below:

After proper-motion correcting the input sample coordinates, we used the GALEXview tool to search within 10\arcsec\,  of each star. This search tool has since been discontinued; however, the data products can still be found on MAST\footnote{\url{https://galex.stsci.edu/GR6/?page=mastform}}. \galex UV data is divided into two bands: near-UV (NUV, 1771 -- 2831 \AA) and far-UV (FUV, 1344 -- 1786 \AA). We utilized only those data which had no bright star window reflection, dichroic reflection, detector run proximity, or bright star ghost photometric flags. For any star that had multiple exposures, we took the weighted mean of the flux density from each individual exposure and the weighted standard deviation as the uncertainty. 

Any star with a magnitude of $<$15 in both the NUV and FUV bands was identified as a lower limit, as this magnitude is where the detector becomes non-linear in response \citep{Morrissey2007}. The lower limits arose only in the K star sample, predominantly in the NUV. Contrary to the analysis method of RY19, we instead use the empirical relations for \galex DR7 data from \citet{Camarota2014} to calculate corrected detections. We then treat these as detections rather than limits in our data. 

Conversely, we calculate upper limits for stars that were observed but not detected by first searching for objects within 10$^{\prime}$ of the stellar coordinates and fitting a power-law to the signal-to-noise ratio (S/N) compared to the signal of each detection within the search radius. We define the upper limit as the magnitude where the $S/N  = 2 $ in this fit. 

\subsection{Subtraction of the Photospheric UV Emission}\label{sec:photsub}
RY19 subtract the photospheric NUV and FUV flux contribution from the K stars in order to determine the evolution of the emission from the chromospheres, transition regions, and coronae of K stars. The photospheric UV fluxes were calculated from PHOENIX stellar atmopshsere models \citep[][]{hauschildt1997, short2005, Hauschildt2006} for stars of 0.6 -- 0.9 M$_{\odot}$. While the photosphere contributes  $\sim10\%$ of the NUV flux for early-type K stars, this decreases to $\sim1\%$ for late-type K stars. The FUV photospheric contribution to the measured flux densities is $<1\%$ in all cases. 

\section{M Star Intrinsic Activity Evolution with Age}\label{sec:evolution}

\subsection{UV Evolution}\label{sec:uvevolution}

We convert the data from the reported \galex magnitudes $m_{GALEX}$ to flux density $F_{GALEX}$ using the following equation from the GALEX FAQ page\footnote{\url{https://asd.gsfc.nasa.gov/archive/galex/FAQ/counts_background.html}}:

\begin{equation}
    F_{GALEX} \textrm{[$\mu$Jy]} = 10^{\frac{23.9-m_{GALEX}}{2.5}}.
\end{equation}

\noindent For consistency, we similarly subtract the photosphere from the M stars as is done in RY19 for K stars, although the contribution is typically less than 1\% in the NUV and negligible in the FUV. With the updated distances from \gaia, we can now analyze the intrinsic UV and X-ray evolutions of M stars and compare these with the intrinsic high-energy evolutions for K stars.

Figure \ref{fig:fluxevolution} shows the NUV and FUV evolutions of K and M stars as a function of age at an absolute distance of 10 pc. Due to a low number of data points at the early ages, we combined the 10 Myr and 24 Myr samples into a single age bin. Since the data contain limits, it is erroneous to simply take the median and direct percentiles of the flux densities. We therefore use a survival function to account for the limits in our data. We fit the Kaplan-Meier estimate for the survival function of the data using the  \texttt{lifelines}\footnote{\url{https://lifelines.readthedocs.io/}} program. From this, we are able to calculate the 50\%, 75\% and 25\% ``survival times" of the data to give us the median and the interquartiles (i.e. 75\% - 25\%: the middle 50\%) of the data, respectively. Due to the different treatment of K star lower limits in this work compared to RY19 (described in \S\ref{sec:photometry}), we are able to apply the Kaplan-Meier estimate to K stars and therefore determine different, more statistically accurate median values than seen in RY19 even though the data remain the same. 

For the M stars, we see that the UV evolution corrected for distance has a similar shape to the distance-independent evolution of SS18. There is a decrease in median flux from 10 to 24 Myr where the stars are still contracting onto the main sequence; however, the interquartiles are still similar out to 150 Myr before declining in both the NUV and FUV bands. The rate of decline is steeper in the FUV than in the NUV, strongly driven by the large number of non-detections for the Hyades-age stars. There is a large (i.e. 2 -- 3 orders of magnitude) spread in the individual age bins that can be seen more clearly in Figures \ref{fig:hist_nuv} and \ref{fig:hist_fuv}. This spread is most likely due to a combination of potential flares in the data, unknown binarity, and different rotation periods within the sample.

We break down the M dwarf sample into late- (0.08 -- 0.35 \ms) and early-type (0.35 -- 0.6 \ms) M dwarfs in order to identify the differences that arise between fully and partially convective stars in Figure \ref{fig:fluxevolution_masses}.  For late M stars, the evolution of both the NUV and FUV flux density maintain a period of saturation out through 150 Myr before declining. Interestingly though, the distance-corrected data show a much earlier decline for early-M stars, with a clear decline between 150 Myr and 650 Myr, which was not seen in the distance-independent analysis in SS18. We discuss the relationship between the distance-independent and distance-corrected approaches and how this leads to different results in \S\ref{sec:comparison}. 

We additionally show the absolute NUV flux density detections compared to the absolute FUV flux density detections in Figure \ref{fig:fuv_v_nuv} to expand Figure 12 of SS18 to the full sample of M stars. Similar to SS18, we see that field-age M stars represent a distinct sample apart from the young moving group and Hyades members, demonstrating the large decline in UV flux between these ages. The best fit line between the early-M stars and late-M stars disagree at field age, demonstrating a larger decline in FUV flux for early-M stars than for late-M stars, which is also seen in the UV/J flux density ratio evolution from SS18.

\subsection{X-ray Evolution}

The X-ray flux is one of the key drivers of ionization of molecules and atoms in planetary atmospheres. 
We further expand on the work of SS18 and SB14 by analyzing the intrinsic X-ray flux of M stars over the whole spectral type, as well as expand the RY19 data to include upper limits for the K star sample. We determine the X-ray flux by searching the Second \rosat All-sky Survey Point Source Catalog (2RXS; 5 -- 124 \AA; \citealt{boller}) for sources within the 3$\sigma$ positional error of 38\arcsec. Using the conversion from hardness ratio and count rate from \citet{schmitt1995}, we calculate the X-ray flux at 10 pc in units of [\fluxcgs]. For any star that was not detected, we follow the same procedure described in \SS\ref{sec:photometry} for calculating the UV upper limits; however, we expand the search to 100\arcmin since 10\arcmin did not include enough detections to make a statistical relationship in order to calculate where $S/N = 2$. 

Figure \ref{fig:xrayevolution} shows the evolution of K, early-M, and late-M star X-ray fluxes with stellar age. For the M stars, we see similar trends as compared to the NUV and FUV, where the flux is roughly constant within the interquartiles for the first 150 Myr before beginning to decline but with a steeper slope than the UV. This agrees with the previous early-M star X-ray study in SB14. We continue to see a large spread in the data, spanning between 1 -- 3 orders of magnitude (Figures \ref{fig:hist_xray_em}, \ref{fig:hist_xray_lm}, and \ref{fig:hist_xray_k}), again most likely due to a combination of rotational differences, unknown binarity, and potential flaring. There are several upper limits at 45 Myr for the K stars weighing down the median of this sample; therefore, while RY19 saw a decrease after 45 Myr, with the inclusion of the upper limits we see instead a saturation period of 150 Myr. This is more consistent with previous X-ray luminosity ($L_X/L_{bol}$) evolution studies for K stars which show a decrease after 80 -- 190 Myr for K stars \citep{jackson12J}.

\subsection{UV vs. X-ray Flux}
Figure \ref{fig:uv_vs_xray} shows the UV flux (see \S\ref{sec:conversions} for flux density to flux conversion methods for low-mass stars) compared to the X-ray flux for detections in both \galex and \rosat. Similar to SB14, there is a main concentration of stars with higher emission and a smaller subset of stars with lower levels of emission. This occurs in both early- and late-type M stars. The low emission group is predominantly comprised of older, field-age stars. For early-M stars, we find similar slopes (0.62 $\pm$ 0.02 and 0.82 $\pm$ 0.03 in the NUV and FUV, respectively) to SB14 for their UV/J versus X-ray/J relationships (0.61 $\pm$ 0.05 and 0.73 $\pm$ 0.06 in the NUV and FUV, respectively). The faster evolution seen in the FUV in Figures \ref{fig:fluxevolution} and \ref{fig:fuv_v_nuv} is also seen slopes of the relationships between NUV and FUV with X-ray, with the FUV slopes being larger than the NUV slopes for both the early- and late-M star cases.

\section{Comparison between M and K star Evolution in the Habitable Zone}\label{sec:hz}

To calculate the habitable zone distances of an example K star (0.8 M$_{\odot}$), early-M star (0.4 M$_{\odot}$), and late-M star (0.1 M$_{\odot}$), we use Equation 3 of \citet{Kopparapu2013}:

\begin{equation}
    d = (\frac{R^2 \cdot T_{\rm eff}^4}{S_{\rm eff}})^{0.5} \text{au},
\end{equation}

where $d$ is the distance from the host star in [au], $R$ and $T_{\rm eff}$ are the stellar radius and effective temperature of the star relative to the solar values, and $S_{\rm eff}$ is the stellar bolometric photospheric flux at the habitable zone. We take the $S_{\rm eff}$ values from \citet{kopparapu14} and interpolate to the appropriate $S_{\rm eff}$ based on the stellar temperatures of our example stars. We finally convert our flux values to the median distance of the edges of the habitable zones of each star (i.e. the middle of the habitable zone). 

Figure \ref{fig:full_nuv_fuv} shows the medians of the K, early-M, and late-type M star absolute UV and X-ray flux from Figures \ref{fig:fluxevolution}, \ref{fig:fluxevolution_masses} and \ref{fig:xrayevolution}. A table version is shown in Table \ref{tab:hz_fluxes}. 

\textit{NUV: }The intrinsic M and K star NUV fluxes show distinct evolutions from each other, with larger masses demonstrating larger fluxes. The NUV for the K star hardly declines, meaning that by field age, the intrinsic flux is over 10 times that of even early-M stars. When converting these absolute fluxes to the centers of the habitable zones of each star, we see that the incident UV fluxes between early-M, late-M, and K stars are overlapping. Since the K stars have farther out habitable zones than M stars, they start out with lower HZ fluxes than the M stars. However, since the K star NUV flux does not decrease significantly while the M stars do have a notable decline, by field age the K stars have twice as much NUV radiation compared to late-M stars and 10 times as much NUV radiation compared to early-M stars. 

\textit{FUV:} The intrinsic M and K star FUV fluxes are more similar to each other than in the NUV case, with the interquartiles overlapping for young and intermediate ages. All three mass ranges appear to decline after 150 Myr, but the rate of decline is shallower for K stars than for either early-M or late-M stars. In the HZ, there is no significant difference between K star and M star FUV flux, although the late-M FUV flux is elevated at 150 Myr compared to both early-Ms and late-Ms. 

\textit{X-ray: }The intrinsic early-M and K star X-ray fluxes are nearly identical at young and intermediate ages, but with K stars again having a shallower decline between 650 Myr to 5 Gyr. Late-M stars start off with a lower intrinsic X-ray flux, but due to their shallower decline than early-M stars, both early- and late-M stars have identical X-ray fluxes by field age. The HZ X-ray fluxes are more separate between K and M stars at young and intermediate ages, with K stars having consistently lower X-ray flux. The more rapid decline of early-M star flux however means that by field age, the HZ X-ray flux is similar between K and early-M stars. The HZ X-ray flux of K stars is constistently between 3 -- 15 times lower than that of late-M stars though. 

Figure \ref{fig:full_nuv_fuv} also suggests that there may be a wavelength dependence on the age of decline for K stars. The M stars appear to consistently decrease after 150 Myr in the NUV, FUV, and X-ray. The K stars on the other hand show hardly any decrease in the NUV through 5 Gyr and a decline after 150 Myr in the FUV and X-ray. Since the X-ray traces the stellar coronae, the FUV traces the transition regions, and the NUV traces the upper chromospheres of stars, this would directly correspond to unique evolutions for different stellar heights, with the age of decline becoming shorter as the distance above the stellar surface increases. Evolutionary observations of various stellar height tracers would help determine if this is the case.

\section{Comparison between Distance-Normalized and Distance-Corrected Methods}\label{sec:comparison}

We consider the differences between the distance-normalized (e.g. UV/J) and distance-corrected (e.g. intrinsic UV flux, corrected to 10 pc) approaches in order to understand the differences between the distance-normalized results presented in SS18 and RY19 and the distance-corrected approach we propose in this study. These relations can be utilized to make more insightful conclusions for future studies without known distances. 

Figure \ref{fig:comps} shows the relationships between the UV flux densities calculated from the distances determined by \gaia to the J-band normalized UV flux values (i.e. distance-independent flux ratios) for K and M stars, respectively. There is relatively small scatter for K dwarfs with an $R^2$ statistic of 0.89 for the FUV data and 0.94 for the NUV data. The spread is wider for M stars with an $R^2$ statistic of 0.65 in the FUV and 0.67 in the NUV. The greater spread for M stars partially arises from the larger luminosity ranges covered by the M star spectral type. 

A variation between the analyses stems from the different slopes of the relationships between the K star flux densities and the M star flux densities. The slopes for the FUV data ($1.08 \pm 0.03$ for K stars and $1.19 \pm 0.05$ for M stars) and NUV data ($1.37 \pm 0.02$ for K stars and $1.47 \pm 0.05$ for M stars) are distinct. The different slopes between K and M stars could lead to a false interpretation of the data as you are no longer comparing the data on the same scales; e.g. a $F_{FUV}/F_J$ result of $10^{-7}$ would be $10^{0}$ $\mu$Jy for K stars and $10^{-1}$ $\mu$Jy for M stars. This makes sense in the context of the results in RY19, who saw similar normalized (UV/J) flux levels between both M and K stars, while in this work using the distance-corrected fluxes we see higher flux levels in K stars compared to M stars. Additionally, because the slopes are $>1$, this would result in the overall trend that the fluxes are less spread out in the $F_{UV}/F_J$ than in the distance-corrected regime, leading to the argument that the fluxes are more similar than they intrinsically are. 

The origin of these differences can be seen in Figure \ref{fig:jbandevolution}, which shows the evolution of the J-band flux for both K and M stars. For K stars, the J-band remains at a constant flux for the entire 5 Gyr, whereas the early-M and late-M stars' J-band evolution declines slightly. This is not an effect of the log-scale plotting -- the p-values of the K, early-M, and late-M stars' median values is 0.96, 0.00, and 0.08 respectively, suggesting that the K star median flux evolution is consistent with a line while both early-M and late-M star median flux evolution is not. While the photosphere does evolve over the stellar lifetime, we would not expect the J-band evolution to be occurring more rapidly than for K stars with shorter lifetimes. This evolution could be a a result of photospheric molecular lines of M stars getting broader as the stars contract onto the main sequence or a greater presence of T Tauri stars that have been shown to exhibit non-photospheric J-band excess \citep[][]{Cieza2005}. A spectroscopic study of M star near-infrared evolution may aid in understanding from where this J-band evolution arises. Additionally, the K star J-band flux is a factor of 10 larger than the M star J-band flux. 

These two factors combined would yield an unrealistic comparison between the distance-normalized K and M star UV and X-ray evolutions since the M star evolution is convolved with the J-band evolution and produces an incorrect shape, while the normalized flux values should not be compared since the J-band fluxes are a factor of 10 separate from each other. 

The distance-independent approach makes for a great substitute for when distances are not available. However, this analysis should be used, when possible, only to compare data sets with similar masses to avoid the issues discussed above. 

% If "omit" error, probably from vspace in the wrong place. Can't submit paper until that's fixed and no red errors. 
\begin{deluxetable*}{r | c c c c c}[t]
\centering
\tablecaption{\textbf{Median Fluxes in the Respective Habitable Zones of K, early-M, and late-M Stars as a Function of Age.} \label{tab:hz_fluxes}}
\tablehead{
\colhead{Stellar Age:}  & \colhead{16.5 Myr} & \colhead{43.5 Myr} & \colhead{150 Myr} & \colhead{650 Myr} & \colhead{5000 Myr}\\
\vspace{-0.55cm}							
}
\startdata			
\multicolumn{6}{c}{NUV HZ Flux [erg s$^{-1}$ cm$^{-2}$]} \\
 % & 16.5 Myr & 43.5 Myr & 150 Myr & 650 Myr & 5000 Myr\\
\hline
K Star	 & 	181	$^{+	166	}_{-	89	}$ & 	118	$^{+	238	}_{-	49	}$ & 	250	$^{+	309	}_{-	11	}$ & 	61	$^{+	137	}_{-	37	}$ & 	40	$^{+	91	}_{-	24	}$ \\
Early-M	 & 	417	$^{+	215	}_{-	154	}$ & 	194	$^{+	31	}_{-	61	}$ & 	141	$^{+	146	}_{-	67	}$ & 	42	$^{+	17	}_{-	17	}$ & 	5.3	$^{+	6.7	}_{-	1.8	}$ \\
Late-M	 & 	612	$^{+	765	}_{-	297	}$ & 	290	$^{+	339	}_{-	134	}$ & 	367	$^{+	332	}_{-	176	}$ & 	79	$^{+	76	}_{-	37	}$ & 	15	$^{+	17	}_{-	11	}$ \\
\hline
\multicolumn{6}{c}{FUV HZ Flux [erg s$^{-1}$ cm$^{-2}$]} \\
\hline
K Star	 & 	26	$^{+	6	}_{-	11	}$ & 	18	$^{+	12	}_{-	10	}$ & 	16	$^{+	50	}_{-	11  }$ & 	3.9	$^{+	7.9	}_{-	3.1	}$ & 	1.2	$^{+	1.3	}_{-	0.4	}$ \\
Early-M	 & 	64	$^{+	78	}_{-	23	}$ & 	38	$^{+	15	}_{-	14	}$ & 	24	$^{+	16	}_{-	3	}$ & 	5.8	$^{+	4.9	}_{-	4.6	}$ & 	0.4	$^{+	0.4	}_{-	0.3	}$ \\
Late-M	 & 	75	$^{+	211	}_{-	40	}$ & 	88	$^{+	82	}_{-	61	}$ & 	123	$^{+	60	}_{-	59	}$ & 	11	$^{+	16	}_{-	2	}$ & 	3.4	$^{+	7.3	}_{-	2.8	}$ \\
\hline
\multicolumn{6}{c}{X-Ray HZ Flux [erg s$^{-1}$ cm$^{-2}$]} \\
\hline		
K Star	 & 	674	$^{+	364	}_{-	252	}$ & 	242	$^{+	592	}_{-	225	}$ & 	244	$^{+	301	}_{-	154	}$ & 	22	$^{+	31	}_{-	17	}$ & 	3.1	$^{+	5.7	}_{-	2.0	}$ \\
Early-M	 & 	1850	$^{+	770	}_{-	300	}$ & 	848	$^{+	448	}_{-	252	}$ & 	710	$^{+	154	}_{-	251	}$ & 	84	$^{+	72	}_{-	0	}$ & 	2.5	$^{+	5.3	}_{-	1.6	}$ \\
Late-M	 & 	2750	$^{+	6380	}_{-	1230	}$ & 	630	$^{+	3690	}_{-	200	}$ & 	2250	$^{+	1550	}_{-	1820	}$ & 	343	$^{+	83	}_{-	98	}$ & 	19	$^{+	129	}_{-	14 }$ \\
\enddata
\end{deluxetable*}

\section{Summary}\label{sec:summary}

We analyzed the evolutions of M star UV and X-ray intrinsic flux as a function of age, building off of the work of \citet{Shkolnik2014} and \citet{Schneider2018} to instead yield distance-corrected flux evolutions with recent \gaia\, EDR3 data. We utilize \galex and \rosat photometry to measure the NUV, FUV, and X-ray fluxes for stars at ages of 10, 24, 45, 150, 650, and 5000 Myr. We compare these results to K star intrinsic fluxes adapted from \citet{richey-yowell2019} and calculated the habitable zone UV and X-ray fluxes of K and M stars. The main results are as follows:

\begin{itemize}
    \item The M star NUV, FUV, and X-ray evolution as a function of age follows a period of constant, saturated activity for $\sim$150 Myr before declining, as is also seen in the fractional flux density evolutions in \citet{Shkolnik2014} and \citet{Schneider2018}.  
    \item Variability among stars of the same age spans from 1 -- 3 magnitudes, most likely due to differences in rotation period, unknown binarity, activity cycles, and flares that may have been caught during the observations.
    \item The FUV and NUV intrinsic fluxes among early-M, late-M, and K stars show distinct evolutions, with K dwarf UV fluxes being 1 -- 2 magnitudes above late-M UV fluxes. 
    \item The FUV and NUV fluxes in the habitable zones of early-M, late-M, and K stars are within the same order of magnitude . 
    \item The K star intrinsic X-ray flux is similar to early-M stars and larger than late-M stars, while late-M dwarfs show 3 -- 15 times larger X-ray fluxes in their HZs than K stars. 
\end{itemize}

Since planets around both K and M stars are receiving large levels of UV radiation, this would infer that their atmospheres are likely undergoing severe photo-dissociation of potential biosignatures or molecules that are necessary for the formation and sustaining of life. However, for planets around K stars with lower X-ray irradiation, there is less heating in the upper atmosphere of the planet, meaning that the planet is more likely to retain an atmosphere. As more young moving groups are identified with future \gaia data, these will refine the early evolution of the high-energy radiation environments provided by stars.

\vspace{0.5cm}
We wish to thank the anonymous referee for a timely and helpful report. T.R.-Y. would like to acknowledge support from the Future Investigators in NASA Earth and Space Exploration (FINESST) award 19-ASTRO20-0081. T.R.-Y. and E.L.S. also acknowledge support from HST-GO-15955.01 from the Space Telescope Science Institute, which is operated by AURA, Inc., under NASA contract NAS 5-26555. 
This work has made use of data from the European Space Agency (ESA) mission
{\it Gaia} (\url{https://www.cosmos.esa.int/gaia}), processed by the {\it Gaia}
Data Processing and Analysis Consortium (DPAC,
\url{https://www.cosmos.esa.int/web/gaia/dpac/consortium}). Funding for the DPAC
has been provided by national institutions, in particular the institutions
participating in the {\it Gaia} Multilateral Agreement. This work is based on observations made with the NASA \textit{Galaxy Evolution Explorer} and the \textit{R{\"O}ntgenSATellit}. \textit{GALEX} is operated for NASA by the California Institute of Technology under NASA contract NAS5-98034. This research utilized the public data from the second ROSAT All-Sky Survey (\url{https://heasarc.gsfc.nasa.gov/W3Browse/rosat/rass2rxs.html}). This work makes use of data products from the Two Micron All-Sky Survey, which is a joint project of the University of Massachusetts and the Infrared Processing and Analysis Center/California Institute of Technology, funded by the National Aeronautics and Space Administration and the National Science Foundation. This research has made use of the SIMBAD database, operated at CDS, Strasbourg, France. 

\newpage
\appendix

\section{Converting UV Flux Densities to Fluxes for Low-Mass Stars}\label{sec:conversions}
We convert the UV data from flux density $F_{GALEX}$ [$\mu$Jy] to flux $f_{GALEX}$ [erg s$^{-1}$ cm$^{-2}$] using

\begin{figure}[t]
    \centering
    \includegraphics[width=\linewidth]{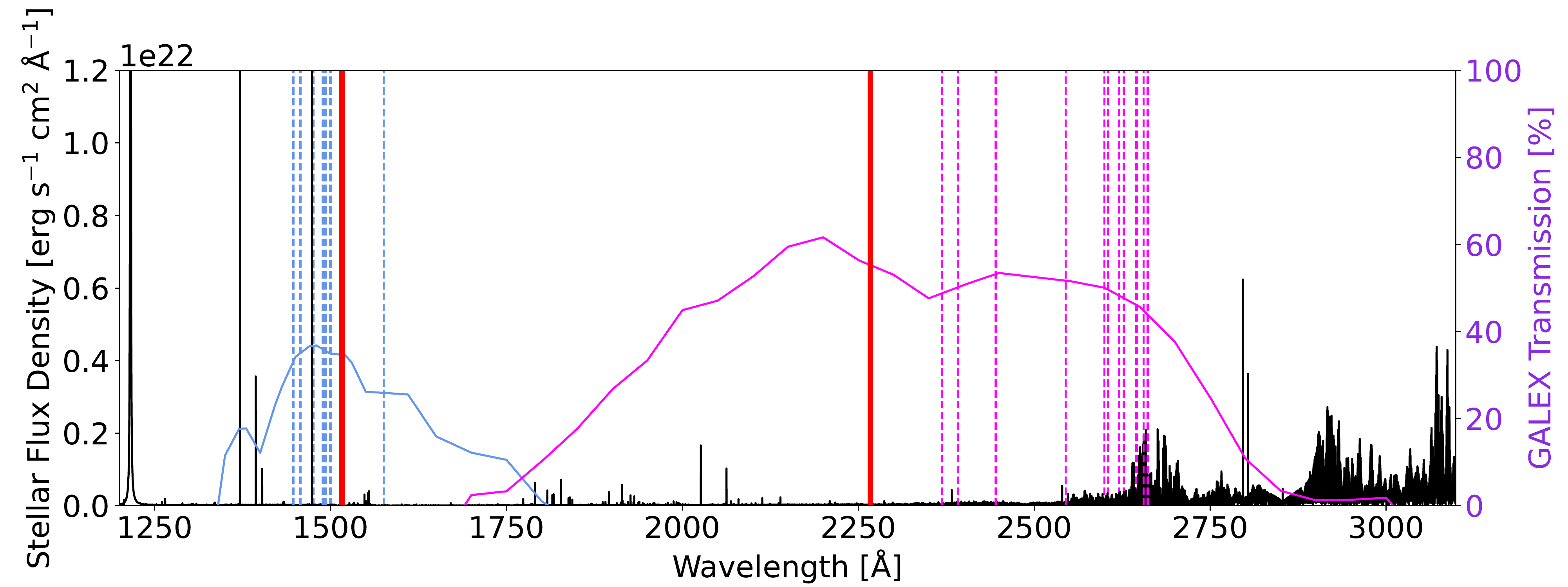}
    \singlespace\caption{Results of the calculation for \galex effective wavelength for K stars. The black shows an example spectrum of a field-age K1 star. Overplotted are the \galex FUV (blue) and NUV (pink) filter transmission profiles. The red lines show the effective wavelengths presented in the \galex documentation for the FUV (left, 1516 \AA) and NUV (right, 2267 \AA). The dotted blue and pink lines show the calculated FUV and NUV effective wavelengths, respectively, for 14 K stars of varying sub-types. The differences between these lines are driven by the spectral energy distribution of each star. }
    \label{fig:effwl}
\end{figure}
% If "omit" error, probably from vspace in the wrong place. Can't submit paper until that's fixed and no red errors. 
\begin{deluxetable}{l c c c}[t]
\centering
\tablecaption{Effective Wavelengths $\lambda_{eff}$ for MUSCLES and HAZMAT K Dwarfs for \galex Filters \label{tab:effwls}}
\tablehead{
\colhead{Star} \vspace{-0.2cm} & \colhead{Spectral} & \colhead{$\lambda_{FUV}$} & \colhead{$\lambda_{ NUV}$}\\
\colhead{Name/Type} & \colhead{Type} & \colhead{[\AA]} & \colhead{[\AA]}\\
\vspace{-0.5cm}							
}
\startdata			
\multicolumn{4}{c}{MUSCLES} \\
\hline
% HD 97658    &   K1  &   1583.5 &   2598.0 \\
Eps Eri     &   K2  &   1575.6 &   2604.6 \\
HD 40307    &   K2.5    &   1500.6 &  2627.2 \\ 
% HD 85512    &   K6  &   1814.1 &   2628.9 \\
\hline
\multicolumn{4}{c}{HAZMAT} \\
\hline
Low Activity  & K0  &   1492.7 & 2620.6\\ 
High Activity & K0 & 1499.1 & 2644.0\\ 
Low Activity & K1 & 1492.3 & 2599.5\\ 
High Activity & K1 &  1475.7 &  2646.6\\
% Low Activity & K2 &  \try{TBD} &\try{TBD}  \\
% High Activity & K2 & \try{TBD}  & \try{TBD} \\
Low Activity & K3 & 1488.8 & 2544.4\\ 
High Activity & K3 & 1456.7&  2655.2  \\
Low Activity & K4 & 1489.6 & 2445.1\\ 
High Activity & K4 & 1457.3 & 2660.7\\ 
Low Activity & K5   & 1492.4 & 2392.2\\
High Activity & K5 &  1449.9 & 2661.3 \\
Low Activity & K6  &  1493.1 &  2368.6\\
High Activity & K6  & 1447.3 &  2655.3 \\
\hline
Median  &   &   1491.0  &   2623.9 \\
\enddata
\end{deluxetable}

\begin{equation}
    f_{GALEX} = (F_{GALEX} \times 10^{-29})c \lambda_{eff}^{-2} \Delta\lambda,
\end{equation}

\noindent where $c$ is the speed of light and the effective wavelength $\lambda_{eff}$ and band width $\Delta\lambda$ change for the FUV and NUV band passes. The band widths for the FUV and NUV are $\Delta\lambda = 268 $\AA and $\Delta\lambda = 732 $\AA\,, respectively. The effective wavelengths reported by \galex are $\lambda_{eff} = 1516 $\AA\, in the FUV and $\lambda_{eff} = 2267 $\AA\, in the NUV. However, these effective wavelengths are calculated on data from Vega and therefore do not account for the spectral shape of M and K stars. 

\citet{Schneider2018} recalculate the effective wavelengths of the \galex band passes for M stars using spectral data of 7 M stars from the MUSCLES Treasury Survey version 2.2 \citep[][]{France2016, Youngblood2016, Loyd2016} ranging from spectral type M1.5 -- M5.5 to determine average $\lambda_{eff}$ values of 1542.8 \AA\, and 2553.4 \AA\, in the FUV and NUV bands, respectively. We adopt these values to convert M star flux density to flux. 

We recalculated the effective wavelength of the GALEX filters using synthetic spectra of K0-6 stars computed with the multi-level non-local thermodynamic equilibrium (non-LTE) code PHOENIX \citep{Hauschildt1993, Hauschildt2006, Baron2007}. We used similar methods to those described in \cite{Peacock2019}, first computing base photosphere models in radiative–convective equilibrium corresponding to the T$_{\rm eff}$, log($g$), and mass of each spectral sub-type. These parameters were determined from \cite{Pecaut2013} and the evolutionary models of \cite{Baraffe2015} for stars aged 5 Gyr. To each of these underlying photospheres, we then superimposed a moderately increasing temperature gradient to simulate a chromosphere until the point at which hydrogen becomes fully ionized, occurring at $\approx$7000 K for K stars. Above the chromosphere, we set a steep temperature gradient to simulate the transition region extending up to 200,000 K, since the majority of UV emission lines form at or below this temperature. For each spectral sub-type, we computed two models with different levels of UV emission, simulated by altering the location and thickness of both the chromosphere and transition region. Each pair of models differs in flux in the GALEX FUV bandpass by a factor of $\sim$50 and in the GALEX NUV bandpass by a factor of $\sim$2. We calculate a median $\lambda_{eff}$ for K dwarfs of 1491.0 \AA\, and 2623.9 \AA\, for the FUV and NUV bands, respectively, as seen in Table \ref{tab:effwls} and Figure \ref{fig:effwl}. Using the \galex reported $\lambda_{effs}$ would underestimate the K star FUV flux by 4\% and overestimate the NUV flux by 34\%.

\bibliography{bibliography}

\end{document}